\documentclass[12pt]{article}

\usepackage{amsmath}
\usepackage{latexsym}

\usepackage{bbm}
\usepackage{epsfig}

\newcommand{\be}{\begin{equation}}
\newcommand{\ee}{\end{equation}}
\newcommand{\ba}{\begin{eqnarray}}
\newcommand{\ea}{\end{eqnarray}}
\newcommand{\no}{\nonumber\\}

\allowdisplaybreaks

\begin{document}

\title{\normalsize \hfill UWThPh-2007-28 \\[8mm]
\LARGE A precision constraint on multi-Higgs-doublet models}

\author{
\\
W.~Grimus,$^{(1)}$\thanks{E-mail: walter.grimus@univie.ac.at}
\ L.~Lavoura,$^{(2)}$\thanks{E-mail: balio@cftp.ist.utl.pt}
\ O.M.~Ogreid,$^{(3)}$\thanks{E-mail: omo@hib.no}
\ and P.~Osland$^{(4)}$\thanks{E-mail: per.osland@ift.uib.no}
\\*[5mm]
$^{(1)} \!$ \small
Fakult\"at f\"ur Physik, Universit\"at Wien \\
\small
Boltzmanngasse 5, 1090 Wien, Austria
\\*[2mm]
$^{(2)} \!$ \small
Universidade T\'ecnica de Lisboa
and Centro de F\'\i sica Te\'orica de Part\'\i culas \\
\small
Instituto Superior T\'ecnico, 1049-001 Lisboa, Portugal
\\*[2mm]
$^{(3)} \!$ \small
Bergen University College, Bergen, Norway
\\*[2mm]
$^{(4)} \!$ \small
Department of Physics and Technology, University of Bergen \\
\small
Postboks 7803, N-5020 Bergen, Norway
\\*[7mm]
}

\date{10 April 2008}

\maketitle

\vspace*{3mm}

\begin{abstract}
We derive a general expression for $\Delta \rho$
(or,
equivalently,
for the oblique parameter $T$)
in the $SU(2) \times U(1)$ electroweak model
with an arbitrary number of scalar $SU(2)$ doublets,
with hypercharge $\pm 1/2$,
and an arbitrary number of scalar $SU(2)$ singlets.
The experimental bound on $\Delta \rho$ constitutes
a strong constraint on the masses and mixings
of the scalar particles in that model.
\end{abstract}

\newpage

\section{Introduction}

In the Standard Model (SM),
the parameter
\be
\label{rho-tree}
\rho = \frac{m_W^2}{m_Z^2 \cos^2{\theta_W}},
\ee
where $m_W$ and $m_Z$ are the masses
of the $W^\pm$ and $Z^0$ gauge bosons,
respectively,
and $\theta_W$ is the weak mixing angle, 
gives the relative strength of the neutral-current
and charged-current interactions
in four-fermion processes at zero momentum transfer~\cite{Ross:1975fq}. 
At tree level $\rho$ is equal to one,
and it remains one even if additional scalar $SU(2)$ doublets,
with hypercharge $\pm 1/2$,
are added to the SM.\footnote{Other scalar $SU(2) \times U(1)$
representations are also allowed,
as long as they have vanishing vacuum expectation values.}
At one-loop level,
the vacuum-polarization effects,
which are sensitive to any field
that couples either to the $W^\pm$ or to the $Z^0$,
produce the vacuum-polarization tensors ($V=W,Z$) 
\be
\label{Eq:vac-pol-tensor}
\Pi_{VV}^{\mu\nu} \left( q \right) =
g^{\mu\nu} A_{VV} \left( q^2 \right)
+ q^\mu q^\nu B_{VV} \left( q^2 \right),
\ee
where $q^\mu$ is the four-momentum of the gauge boson.
Then,
deviations of $\rho$ from unity arise,
which are determined
by the self-energy difference~\cite{Ross:1975fq,Einhorn:1981cy}
\be
\label{Eq:delta-rho-def}
\frac{A_{WW} \left( 0 \right)}{m_W^2}
- 
\frac{A_{ZZ} \left( 0 \right)}{m_Z^2}.
\ee
The precise measurement~\cite{Yao:2006px},
at LEP,
of the
$W^\pm$ and $Z^0$ self-energies
is in striking agreement with the SM predictions~\cite{delta-rho-th} and
provides a strong constraint on extended electroweak models.
For instance,
one can constrain the two-Higgs-doublet model (2HDM)
in this way~\cite{bertolini, denner}.

In this paper we are interested
in the contributions to the $\rho$ parameter
generated by an extension of the SM.
Therefore,
we define a $\Delta\rho$ which refers
to the non-SM part of the quantity~(\ref{Eq:delta-rho-def}):
\be
\label{delta-rho0}
\Delta \rho = 
\left[
\frac{A_{WW} \left( 0 \right)}{m_W^2}
-
\frac{A_{ZZ} \left( 0 \right)}{m_Z^2}
\right]_\mathrm{SM\ extension}
- 
\left[
\frac{A_{WW} \left( 0 \right)}{m_W^2}
-
\frac{A_{ZZ} \left( 0 \right)}{m_Z^2}
\right]_\mathrm{SM}.
\ee
The SM contributions to the quantity~(\ref{Eq:delta-rho-def})
are known up to the leading terms at three-loop level~\cite{bij}.
However,
the consistent SM subtraction in equation~(\ref{delta-rho0})
only requires the
one-loop SM result.
In the same vein,
we are allowed to make the replacement
$m_Z^2 = m_W^2 / c_W^2$ in equation~(\ref{delta-rho0}),
writing instead
\be
\label{subtract}
\Delta \rho = 
\left[
\frac{A_{WW} \left( 0 \right) - c_W^2 A_{ZZ} \left( 0 \right)}{m_W^2}
\right]_\mathrm{SM\ extension}
- 
\left[
\frac{A_{WW} \left( 0 \right) - c_W^2 A_{ZZ} \left( 0 \right)}{m_W^2}
\right]_\mathrm{SM}.
\ee
Here and in the following, 
we use the abbreviations 
$c_W = \cos{\theta_W}$,
$s_W =\sin{\theta_W}$. 

At one loop,
the contributions of new physics to the self-energies
constitute intrinsically divergent Feynman diagrams,
but the divergent parts cancel out among different diagrams, 
between $A_{WW} \left( 0 \right)$
and $c_W^2 A_{ZZ} \left( 0 \right)$,
and also,
eventually, 
through the subtraction of the SM contributions
laid out in equation~(\ref{subtract}).
If the new-physics model is renormalizable,
then $\Delta \rho$ is finite.
The cancellations finally leave
either a quadratic or a logarithmic dependence
of $\Delta \rho$ on the masses of the new-physics particles.
The pronounced effects of large masses
is what renders the parameter $\Delta\rho$
so interesting for probing physics beyond the Standard Model.

The functions $A_{VV} \left( q^2 \right)$
contain more information about new physics
than the one just provided by $\Delta \rho$.
In fact,
for new physics much above the electroweak scale, 
a detailed analysis of the so-called ``oblique corrections''
lead to the identification of three relevant 
observables, which were called $S$,
$T$ and $U$ in~\cite{Peskin:1990zt} and $\epsilon_1$,
$\epsilon_2$ and $\epsilon_3$
in~\cite{Altarelli:1990zd}.\footnote{For new
physics at a mass scale comparable to the electroweak 
scale three more such ``oblique parameters'' have been identified
in~\cite{maksymyk}.} 
While these two sets of observables differ in their precise definitions,
the quantity of interest in this paper is simply
\be
\Delta\rho = \alpha T = \epsilon_1,
\ee
where
$\alpha = e^2 / \left( 4 \pi \right) = g^2 s_W^2 / \left( 4 \pi \right)$
is the fine-structure constant.

It is not straightforward to obtain a bound on $\Delta \rho$
from electroweak precision data.
One possibility is to add the oblique parameters
to the SM parameter set and perform fits to the data.
However,
since the SM Higgs-boson loops themselves resemble oblique effects,
one cannot determine the SM Higgs-boson mass $m_h$
simultaneously with $S$ and $T$~\cite{delta-rho-th}.
To get a feeling for the order of magnitude allowed for $\Delta \rho$,
we quote the number 
\be
\label{value}
T = -0.03 \pm 0.09\; (+0.09),
\ee
which was obtained in~\cite{delta-rho-th} by fixing $U=0$.
For the mean value of $T$,
the Higgs-boson mass $m_h = 117$ GeV was assumed;
the mean value in parentheses is for $m_h = 300$ GeV.
Equation~(\ref{value}) translates into 
$\Delta \rho = -0.0002 \pm 0.0007\; (+0.0007)$.

There is a vast literature on the
2HDM---see~\cite{higgs-hunter} for a review,
\cite{hollik} for the renormalization of the model,
\cite{chankowski1,chankowski2} for the possibility of having
a light pseudoscalar compatible with all experimental constraints,
and~\cite{kane,osland},
and the references therein,
for other various recent works.
However,
just as the 2HDM may differ significantly from the SM, 
a general multi-Higgs-doublet model may be quite different
from its minimal version with only two Higgs doublets~\cite{grossman}. 
Three or more Higgs doublets
frequently appear in models with family symmetries
through which one wants to explain various features
of the fermion masses and mixings;
for some examples in the lepton sector see the reviews in~\cite{models}.

In this paper we present a calculation
of $\Delta \rho$
in an extension of the SM
with an arbitrary number of Higgs doublets and also,
in addition,
arbitrary numbers of neutral and charged scalar $SU(2)$ singlets. 
Our results can be used to check
the compatibility of the scalar sector of multi-Higgs models
with the constraints resulting from the electroweak precision experiments. 

Recently,
there has been some interest
in ``dark'' scalars~\cite{Ma:2006km,Barbieri:2006dq}.
These are scalars that have no Yukawa couplings,
and are thus decoupled from ordinary matter.
Furthermore,
they have no vacuum expectation values (VEVs)
and therefore display truncated couplings to the gauge bosons.
However,
they would have quadrilinear vector--vector--scalar--scalar
and trilinear vector--scalar--scalar
(but no vector--vector--scalar) couplings,
and would thus also contribute to,
and be constrained by,
$\Delta\rho$.

The plan of the paper is as follows.
In Section~2 we present a description of our extension of the SM
and the final result of the calculation of $\Delta \rho$;
this section is self-consistent
and the result can be used without need to consult the rest of the paper.
The details of the calculation are laid out in Section~3.
The application of our $\Delta \rho$ formula
to the general 2HDM is given in Section~4.
The summary of our study is found in Section~5.

\section{The model and the result for $\Delta \rho$}

\subsection{The model}

We consider an $SU(2) \times U(1)$ electroweak model
in which the scalar sector includes $n_d$ $SU(2)$ doublets
with hypercharge $1/2$,\footnote{Equivalently,
we may consider the model to contain
$SU(2)$ doublets with hypercharge $-1/2$,
since
\[
\tilde \phi_k \equiv i \tau_2 \phi_k^\ast = \left( \begin{array}{c}
{\varphi_k^0}^\ast \\ - \varphi_k^- \end{array} \right)
\]
is also a doublet of $SU(2)$.}
\be
\phi_k = \left( \begin{array}{c}
\varphi_k^+ \\ \varphi_k^0 \end{array} \right),
\quad k = 1, 2, \ldots, n_d.
\ee
Moreover,
we allow the model to include an arbitrary number and variety
of $SU(2)$-singlet scalars;
in particular,
$n_c$ complex $SU(2)$ singlets with hypercharge $1$,
\be
\chi_j^+, \quad j = 1, 2, \ldots, n_c
\ee
and $n_n$ real $SU(2)$ singlets with hypercharge $0$,
\be
\chi_l^0, \quad l = 1, 2, \ldots, n_n.
\ee
In general,
our model may include other scalar fields,
singlet under the gauge $SU(2)$,
with different electric charges.

The neutral fields are allowed to have vacuum expectation values (VEVs).
Thus,
\ba
\left\langle 0 \left| \varphi_k^0 \right| 0 \right\rangle &=&
\frac{v_k}{\sqrt{2}},
\\
\left\langle 0 \left| \chi_l^0 \right| 0 \right\rangle &=& u_l,
\ea
the $v_k$ being in general complex.
(The $u_l$ are real since the $\chi_l^0$ are real fields.)
We define as usual
$v = \left( \sum_{k=1}^{n_d} \left| v_k \right|^2 \right)^{1/2}
\simeq 246\, \mbox{GeV}$.
Then,
the masses of the $W^\pm$ and $Z^0$ gauge bosons are,
at tree level,
$m_W = g v / 2$ and $m_Z = m_W / c_W$,
respectively.\footnote{Since the neutral singlet fields
carry no hypercharge,
their VEVs $u_l$ do not contribute to the masses of the gauge bosons.}
We expand the neutral fields around their VEVs,
\ba
\varphi_k^0 &=& \frac{1}{\sqrt{2}} \left(v_k+ \varphi_k^0{}^\prime\right),
\\
\chi_l^0 &=& u_l + \chi_l^0{}^\prime.
\ea

Altogether,
there are $n = n_d + n_c$ complex scalar fields
with electric charge $1$
and $m = 2 n_d + n_n$ real scalar fields with electric charge $0$.
The mass matrices of all these scalar fields
will in general lead to their mixing.
The physical (mass-eigenstate) charged and neutral scalar fields
will be called $S_a^+$ ($a = 1, 2, \ldots, n$)
and $S_b^0$ ($b = 1, 2, \ldots, m$),
respectively.
Note that the fields $S_b^0$ are real.
We use $m_a$ to denote the mass of $S_a^\pm$
and $\mu_b$ to denote the mass of $S_b^0$.
We have
\ba
\varphi_k^+ &=& \sum_{a=1}^n U_{ka} S_a^+,
\label{Umatrix} \\
\chi_j^+ &=& \sum_{a=1}^n T_{ja} S_a^+,
\\
\varphi_k^0{}^\prime &=& \sum_{b=1}^m V_{kb} S^0_b,
\\
\chi_l^0{}^\prime &=& \sum_{b=1}^m R_{lb} S^0_b,
\ea
the matrices $U$,
$T$,
$V$ and $R$ having dimensions $n_d \times n$,
$n_c \times n$,
$n_d \times m$ and $n_n \times m$,
respectively.
The matrix $R$ is real,
the other three are complex.
The matrix
\be
\tilde U \equiv \left( \begin{array}{c} U \\ T \end{array} \right)
\label{tildeU}
\ee
is $n \times n$ unitary;
it is the matrix which diagonalizes the (Hermitian) mass matrix
of the charged scalars.
The real matrix
\be
\tilde V \equiv \left( \begin{array}{c}
\mbox{Re}\, V \\ \mbox{Im}\, V \\ R \end{array} \right)
\label{tildeV}
\ee
is $m \times m$ orthogonal;
it diagonalizes the (symmetric) mass matrix
of the real components of the neutral-scalar fields.\footnote{Our
treatment of the mixing of scalars is inspired by~\cite{grimus}.}

There are in the spontaneously broken $SU(2) \times U(1)$ theory
three unphysical Goldstone bosons,
$G^\pm$ and $G^0$.
For definiteness we assign to them the indices $a=1$ and $b=1$,
respectively:
\ba
S_1^\pm &=& G^\pm,
\\
S_1^0 &=& G^0.
\ea
Thus,
only the $S_a^\pm$ with $a \ge 2$ are physical and,
similarly,
only the $S_b^0$ with $b \ge 2$ correspond to true particles.
In the general 't~Hooft gauge that we shall use in our computation,
the masses of $G^\pm$ and $G^0$ are arbitrary and unphysical,
and they cannot appear in the final result for $\Delta \rho$.

\subsection{The result}

As we shall demonstrate in the next section,
the value of $\Delta \rho$ in the model outlined above is
\begin{subequations}
\label{final}
\ba
\Delta \rho &=& \frac{g^2}{64 \pi^2 m_W^2} \left\{
\sum_{a=2}^n\, \sum_{b=2}^m\,
\left| \left( U^\dagger V \right)_{ab} \right|^2
F \left( m_a^2, \mu_b^2 \right)
\right. \label{1a} \\ & &
- \sum_{b=2}^{m-1}\, \sum_{b^\prime = b+1}^m\,
\left[ \mbox{Im} \left( V^\dagger V \right)_{b b^\prime} \right]^2
F \left( \mu_b^2, \mu_{b^\prime}^2 \right)
\label{1b} \\ & &
- 2\, \sum_{a=2}^{n-1}\, \sum_{a^\prime = a+1}^n\,
\left| \left( U^\dagger U \right)_{a a^\prime} \right|^2
F \left( m_a^2 , m_{a^\prime}^2 \right)
\label{1c} \\ & &
+ 3\, \sum_{b=2}^m\,
\left[ \mbox{Im} \left( V^\dagger V \right)_{1b} \right]^2
\left[
F \left( m_Z^2, \mu_b^2 \right) - F \left( m_W^2, \mu_b^2 \right)
\right]
\label{1d} \\ & & \left.
- 3 \left[
F \left( m_Z^2, m_h^2 \right) - F \left( m_W^2, m_h^2 \right)
\right]
\right\},
\label{1e}
\ea
\end{subequations}
where $m_a$, $m_{a^\prime}$
denote the masses of the charged scalars
and $\mu_b$, $\mu_{b^\prime}$
denote the masses of the neutral scalars.
The term~(\ref{1b}) contains a sum over all pairs
of {\em different} physical neutral scalar particles
$S_b^0$ and $S_{b^\prime}^0$;
similarly,
the term~(\ref{1c}) contains a sum over all
pairs of different charged scalars,
excluding the Goldstone bosons $G^\pm$,
i.e.~$2 \le a < a^\prime \le n$.
The term~(\ref{1e}) consists of the subtraction,
from the rest of $\Delta \rho$,
of the SM result---$m_h$ is the mass
of the sole SM physical neutral scalar,
the so-called Higgs particle.

In equation~(\ref{final}),
the function $F$ of two non-negative arguments $x$ and $y$ is
\be \label{Eq:F}
F \left( x, y \right) \equiv
\left\{ \begin{array}{ll}
{\displaystyle \frac{x+y}{2} - \frac{xy}{x-y}\, \ln{\frac{x}{y}}}
&\Leftarrow\ x \neq y,
\\*[3mm]
0 &\Leftarrow\ x = y.
\end{array} \right.
\ee
This is a non-negative function,
symmetrical under the interchange of its two arguments,
and vanishing if and only if those two arguments are equal.
This function has the important property
that it grows linearly with $\max(x,y)$,
i.e.~quadratically with the heaviest-scalar mass,
when that mass becomes very large.
Unless there are cancellations,
this leads to a quadratic divergence of $\Delta\rho$
for very heavy scalars (Higgs bosons).

If there are in the model any $SU(2)$-singlet scalars
with electric charge other than $0$ or $\pm 1$,
then the existence of those scalars does not contribute to $\Delta \rho$,
they do not modify equation~(\ref{final}),
at one-loop level,
in any way.

A simplification occurs when there are in the model
no $SU(2)$-singlet charged scalars $\chi_j^+$.
In that case,
there is no matrix $T$,
hence the matrix $U$ is unitary by itself,
and the term~(\ref{1c}) vanishes.

When there are in the model no $SU(2)$-singlet neutral scalars $\chi_l^0$,
there is no matrix $R$,
hence $\mbox{Re} \left( V^\dagger V \right)_{b b^\prime}
= \left( \mbox{Re}\, V^T\ \mbox{Re}\, V
+ \mbox{Im}\, V^T\ \mbox{Im}\, V \right)_{b b^\prime}
= \delta_{b b^\prime}$.
Then,
in the terms~(\ref{1b}) and~(\ref{1d}) one may write
$\left| \left( V^\dagger V \right)_{b b^\prime} \right|^2$
instead of
$\left[ \mbox{Im} \left( V^\dagger V \right)_{b b^\prime} \right]^2$.

Thus,
in an $n_d$-Higgs-doublet model \emph{without any scalar singlets},
one has simply
\ba
\Delta \rho &=& \frac{g^2}{64 \pi^2 m_W^2} \left\{
\sum_{a=2}^{n_d}\, \sum_{b=2}^{2 n_d}\,
\left| \left( U^\dagger V \right)_{ab} \right|^2
F \left( m_a^2, \mu_b^2 \right)
\right. \no & &
- \sum_{b=2}^{2 n_d -1}\, \sum_{b^\prime = b+1}^{2 n_d}\,
\left| \left( V^\dagger V \right)_{b b^\prime} \right|^2
F \left( \mu_b^2, \mu_{b^\prime}^2 \right)
\no & &
+ 3\, \sum_{b=2}^{2 n_d}\, 
\left| \left( V^\dagger V \right)_{1b} \right|^2
\left[
F \left( m_Z^2, \mu_b^2 \right) - F \left( m_W^2, \mu_b^2 \right)
\right]
\no & & \left.
- 3 \left[
F \left( m_Z^2, m_h^2 \right) - F \left( m_W^2, m_h^2 \right)
\right]
\right\}.
\label{finalMHDM}
\ea
Our general results have been checked to be consistent with 
specific results for $\Delta\rho$ in a few models. 
These include the results for both the CP 
conserving version~\cite{bertolini,chankowski1} 
and the CP non-conserving version~\cite{osland} of the 
2HDM.\footnote{There is some discrepancy between our result and the
  one presented in Section~4 of~\cite{higgs-hunter}.}
It has also been checked against a model containing one doublet 
and one scalar singlet \cite{profumo}.

\section{Derivation of the result}

This section contains the derivation of equation~(\ref{final}).
It may be skipped by those who are not interested
in the details of that derivation.

\subsection{The Lagrangian}

We use the conventions of~\cite{book}.
The covariant derivative of the doublets is
\be
D_\mu \phi_k = \left( \begin{array}{c}
\partial_\mu \varphi_k^+
- i\, {\displaystyle \frac{g}{\sqrt{2}}}\, W_\mu^+ \varphi_k^0
+ i\, {\displaystyle \frac{g \left( s_W^2 - c_W^2 \right)}
{2 c_W}}\,
Z_\mu \varphi_k^+
+ i e A_\mu \varphi_k^+
\\
\partial_\mu \varphi_k^0
- i\, {\displaystyle \frac{g}{\sqrt{2}}}\, W_\mu^- \varphi_k^+
+ i\, {\displaystyle \frac{g}{2 c_W}}\, Z_\mu \varphi_k^0
\end{array} \right)
\ee
and the covariant derivative of the charged singlets is
\be
D_\mu \chi_j^+ = \partial_\mu \chi_j^+
+ i\, \frac{g s_W^2}{c_W}\, Z_\mu \chi_j^+
+ i e A_\mu \chi_j^+.
\ee
The covariant derivative of the neutral singlets is,
of course,
just identical with their ordinary derivative.
We use the unitarity of $\tilde U$ in equation~(\ref{tildeU}),
in particular
\be
\left( T^\dagger T \right)_{a^\prime a} = \delta_{a^\prime a}
- \left( U^\dagger U \right)_{a^\prime a}.
\ee
We also use the orthogonality of $\tilde V$ in equation~(\ref{tildeV})
to arrive at the gauge-kinetic Lagrangian
\begin{subequations}
\ba
& & \sum_{k=1}^{n_d}
\left( D^\mu \phi_k \right)^\dagger
\left( D_\mu \phi_k \right)
+ \sum_{j=1}^{n_c}
\left( D^\mu \chi_j^- \right)
\left( D_\mu \chi_j^+ \right)
+ \frac{1}{2}\, \sum_{l = 1}^{n_n}
\left( \partial^\mu \chi_l^0  \right)
\left( \partial_\mu \chi_l^0 \right)
\no &=&
\sum_{a=1}^n
\left( \partial^\mu S_a^- \right)
\left( \partial_\mu S_a^+ \right)
+ \frac{1}{2}\, \sum_{b=1}^m
\left( \partial^\mu S_b^0 \right)
\left( \partial_\mu S_b^0 \right)
\\ & &
+ m_W^2 W^{\mu -} W_\mu^+
+ m_Z^2\, \frac{Z^\mu Z_\mu}{2}
\\ & &
+ i m_W \sum_{a=1}^n
\left[ W_\mu^- \left( \omega^\dagger U \right)_a \partial^\mu S_a^+
- W_\mu^+ \left( U^\dagger \omega \right)_a \partial^\mu S_a^- \right]
\label{Goldcharged} \\ & &
+ m_Z Z_\mu \sum_{b=1}^m\, \mbox{Im} \left( \omega^\dagger V \right)_b
\partial^\mu S_b^0
\label{Goldneutral} \\ & &
- \left( e m_W A^\mu + g s_W^2 m_Z Z^\mu \right)
\sum_{a=1}^n \left[ \left( \omega^\dagger U \right)_a W_\mu^- S_a^+
+ \left( U^\dagger \omega \right)_a W_\mu^+ S_a^- \right]
\label{Gold1} \\ & &
+ i e A_\mu \sum_{a=1}^n \left( S_a^+ \partial^\mu S_a^-
- S_a^- \partial^\mu S_a^+ \right)
\\ & &
+ i\, \frac{g}{2 c_W}\, Z_\mu
\sum_{a, a^\prime = 1}^n
\left[ 2 s_W^2 \delta_{a a^\prime}
- \left( U^\dagger U \right)_{a^\prime a} \right]
\left( S_a^+ \partial^\mu S_{a^\prime}^-
- S_{a^\prime}^- \partial^\mu S_a^+ \right)
\label{int8} \\ & &
+ \frac{g}{2 c_W}\, Z_\mu\,
\sum_{b=1}^{m-1} \sum_{b^\prime = b+1}^{m}\,
\mbox{Im} \left( V^\dagger V \right)_{b b^\prime}
\left( S_b^0 \partial^\mu S_{b^\prime}^0
- S_{b^\prime}^0 \partial^\mu S_b^0 \right)
\label{int7} \\ & &
+ i\, \frac{g}{2}\, \sum_{a=1}^n \sum_{b=1}^m
\left[ \left( U^\dagger V \right)_{ab} W_\mu^+
\left( S_a^- \partial^\mu S_b^0 - S_b^0 \partial^\mu S_a^- \right)
\right. \no & & \left.
+ \left( V^\dagger U \right)_{ba} W_\mu^-
\left( S_b^0 \partial^\mu S_a^+ - S_a^+ \partial^\mu S_b^0 \right)
\right]
\label{int6} \\ & &
+ g \left( m_W W_\mu^+ W^{\mu -}
+ \frac{m_Z}{c_W}\, \frac{Z_\mu Z^\mu}{2} \right)
\sum_{b=1}^m S_b^0\, \mbox{Re} \left( \omega^\dagger V \right)_b
\label{int1} \\ & &
- \left( \frac{eg}{2}\, A^\mu + \frac{g^2 s_W^2}{2 c_W}\, 
Z^\mu \right)
\sum_{a=1}^n \sum_{b=1}^m S_b^0
\left[ \left( U^\dagger V \right)_{ab} W_\mu^+ S_a^-
+ \left( V^\dagger U \right)_{ba} W_\mu^- S_a^+ \right]
\\ & &
+ \left( \frac{g^2}{4}\, W^{\mu -} W_\mu^+
+ \frac{g^2}{4 c_W^2}\, \frac{Z^\mu Z_\mu}{2} \right)
\sum_{b, b^\prime = 1}^m
\left( V^\dagger V \right)_{b^\prime b} S^0_{b^\prime} S^0_b
\label{int2} \\ & &
+ \frac{g^2}{2}\, W^{\mu -} W_\mu^+
\sum_{a, a^\prime = 1}^n
\left( U^\dagger U \right)_{a^\prime a} S_{a^\prime}^- S_a^+
\label{int3} \\ & &
+ 2 e^2\, \frac{A^\mu A_\mu}{2}\, \sum_{a=1}^n S_a^- S_a^+
\\ & &
+ \frac{e g}{c_W}\, A^\mu Z_\mu
\sum_{a, a^\prime = 1}^n
\left[ 2 s_W^2 \delta_{a a^\prime}
- \left( U^\dagger U \right)_{a^\prime a} \right]
S_{a^\prime}^- S_a^+
\\ & &
+ \frac{g^2}{2 c_W^2}\, \frac{Z^\mu Z_\mu}{2}\,
\sum_{a, a^\prime = 1}^n
\left[ 4 s_W^4 \delta_{a a^\prime} + \left( 1 - 4 s_W^2 \right)
\left( U^\dagger U \right)_{a^\prime a} \right]
S_{a^\prime}^- S_a^+.
\label{int4}
\ea
\end{subequations}
In lines~(\ref{Goldcharged})--(\ref{Gold1}) and~(\ref{int1})
we have used an $n_d$-vector $\omega$
defined by $\omega_k \equiv v_k / v$.
By identifying lines~(\ref{Goldcharged}) and~(\ref{Goldneutral})
with the usual terms~\cite{book}
mixing the $W^\pm$ and $Z^0$ gauge bosons
with the $G^\pm$ and $G^0$ Goldstone bosons,
respectively,
\[
i m_W \left( W_\mu^- \partial^\mu G^+ - W_\mu^+ \partial^\mu G^- \right)
+ m_Z Z_\mu \partial^\mu G^0,
\]
we conclude that the components of the Goldstone bosons
are given by~\cite{grimus} 
\ba
& & U_{k1} = \frac{v_k}{v}, \quad \mbox{hence} \quad T_{j1} = 0,
\label{uk1} \\
& & V_{k1} = i\, \frac{v_k}{v}, \quad \mbox{hence} \quad R_{l1} = 0.
\label{vk1}
\ea
Therefore,
we may rewrite line~(\ref{Gold1}) as
\be
- \left( e m_W A^\mu + g s_W^2 m_Z Z^\mu \right)
\left( W_\mu^- G^+ + W_\mu^+ G^- \right)
\ee
and line~(\ref{int1}) as
\be
- g \left( m_W W_\mu^+ W^{\mu -}
+ \frac{m_Z}{c_W}\, \frac{Z_\mu Z^\mu}{2} \right)
\sum_{b=2}^m S_b^0\, \mbox{Im} \left( V^\dagger V \right)_{1b}.
\label{proviso}
\ee
The sum starts at $b=2$
because $\mbox{Im} \left( V^\dagger V \right)_{11} = 0$.

If there are in the theory any $SU(2)$-singlet scalars $S^{\pm Q}$
with electric charge $\pm Q$ other than $0$ or $\pm 1$,
then those scalars do not mix with components of the doublets.
Their covariant derivative is
\be
D_\mu S^{+Q} = \partial_\mu S^{+Q}
+ i\, \frac{g s_W^2 Q}{c_W}\, Z_\mu S^{+Q}
+ i e Q A_\mu S^{+Q}.
\ee
This yields,
in particular,
the following two interaction terms in the Lagrangian:
\begin{subequations}
\label{Lag}
\ba
& \mathcal{L} = \cdots \! \!& +i\, \frac{g s_W^2 Q}{c_W}\, Z_\mu 
\left( S^{+Q} \partial^\mu S^{-Q}  
- S^{-Q} \partial^\mu S^{+Q} \right)
\label{Lag1} \\ & &
+ \left( \frac{g s_W^2 Q}{c_W} \right)^2 
Z_\mu Z^\mu S^{-Q} S^{+Q}.
\label{Lag2}
\ea
\end{subequations}

\subsection{The Feynman diagrams}
\begin{figure}
\begin{center}
\begin{tabular}{ccc}
\epsfig{file=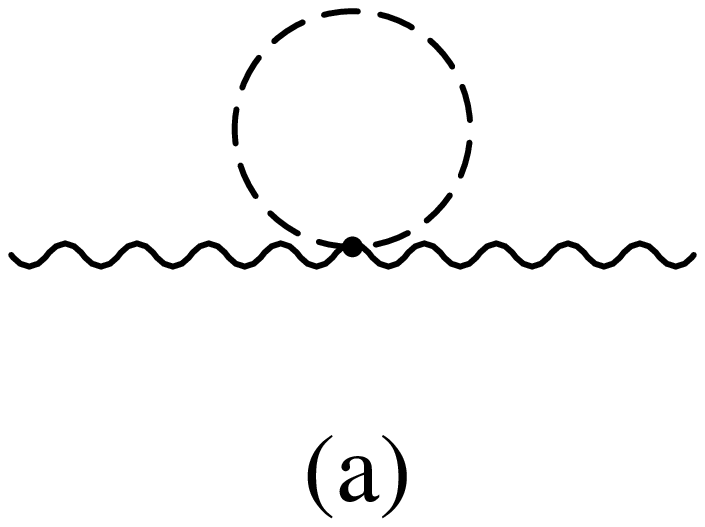,width=3.5cm} &
\epsfig{file=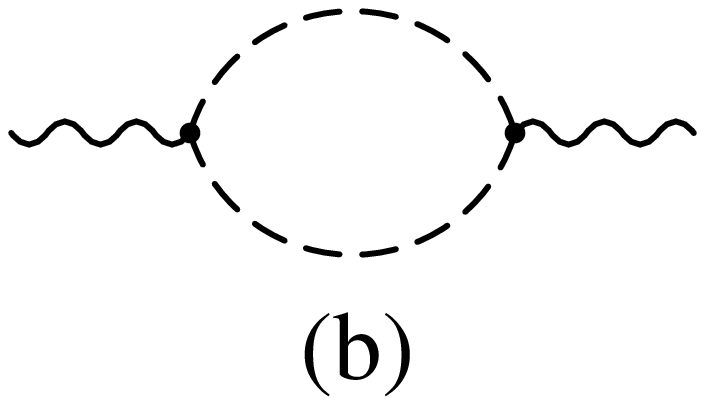,width=3.5cm} &
\epsfig{file=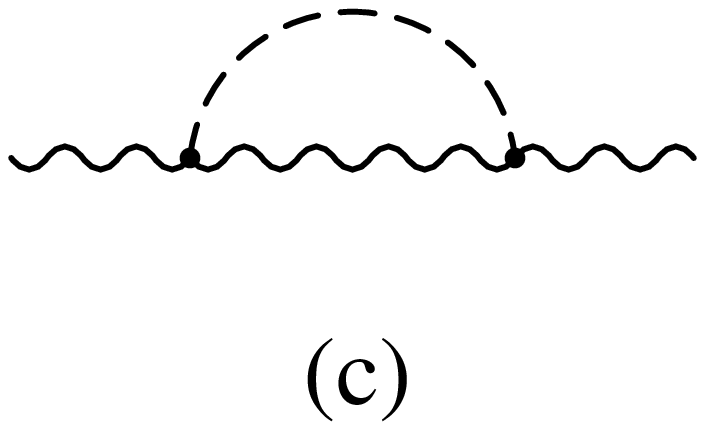,width=3.5cm} 
\end{tabular}
\end{center}
\caption{Three types of Feynman diagrams
occurring in the calculation of the vacuum polarizations.
\label{fig}}
\end{figure}
\begin{figure}
\begin{center}
\begin{tabular}{ccc}
\epsfig{file=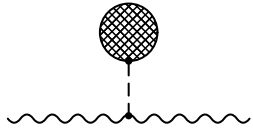,width=3.5cm}
\end{tabular}
\end{center}
\caption{Tadpole diagrams which do not contribute to $\Delta \rho$.
\label{fig-tadpoles}}
\end{figure}
In our model,
in the computation of the vacuum polarizations
of the gauge bosons $W^\pm$ and $Z^0$
there are four types of Feynman diagrams involving scalar fields:
\begin{description}
\item {\it Type (a) diagrams}:
A scalar branches off from the gauge-boson line
and loops back to \emph{the same point} in that gauge-boson line---see
figure~\ref{fig}(a).
When the scalar is neutral,
the relevant interaction terms in the Lagrangian
are the ones in line~(\ref{int2}),
for $b^\prime = b$;
but then the contribution to $\Delta \rho$ vanishes,
since one obtains $\Pi^{\mu \nu}_{WW} = c_W^2 \Pi^{\mu \nu}_{ZZ}$.
When the scalar is charged,
the relevant terms in the Lagrangian are those
in line~(\ref{int3}) for $\Pi^{\mu \nu}_{WW}$
and line~(\ref{int4}) for $\Pi^{\mu \nu}_{ZZ}$,
in both cases for $a^\prime = a$.
\item {\it Type (b) diagrams}:
The gauge-boson line splits into two scalar lines
which later reunite to form a new gauge-boson line---see
figure~\ref{fig}(b).
The relevant terms in the Lagrangian are
those in line~(\ref{int6}) for $\Pi^{\mu \nu}_{WW}$,
and those in lines~(\ref{int8}) and (\ref{int7})
for $\Pi^{\mu \nu}_{ZZ}$.
\item {\it Type (c) diagrams}:
A neutral scalar branches off from the gauge-boson line
and loops to \emph{a later point} in that gauge-boson line---see
figure~\ref{fig}(c).
The interaction terms in the Lagrangian
responsible for these Feynman diagrams are those in expression~(\ref{proviso}).
\item {\it Type (d) diagrams}:
A neutral scalar branches off,
with zero momentum,
from the gauge-boson line,
and produces a loop of some stuff---see figure~\ref{fig-tadpoles}.
These ``tadpole'' Feynman diagrams
originate from the interaction terms in expression~(\ref{proviso}).
They yield a vanishing contribution to $\Delta \rho$ since
one obtains $\Pi^{\mu \nu}_{WW} = c_W^2 \Pi^{\mu \nu}_{ZZ}$.
Hence we may omit the tadpole diagrams altogether.
\end{description}

\subsection{Computation of the loop diagrams}

We use dimensional regularization
in the computation of the Feynman diagrams.
The dimension of space--time is $d$.
An unphysical mass $\mu$ is used
to keep the dimension of each integral unchanged
when $d$ varies.
We define the divergent quantity
\[
\mbox{div} \equiv \frac{2}{4-d} - \gamma + 1
+ \ln{\left( 4 \pi \mu^2 \right)},
\]
where $\gamma$ is Euler's constant.
In the computation of type (a) Feynman diagrams
the relevant momentum integral is
\be
\mu^{4-d} \int \frac{\mbox{d}^d k}{\left( 2 \pi \right)^d}\,
\frac{g^{\mu \nu}}{k^2 - A + i \varepsilon}
= \frac{i g^{\mu \nu}}{16 \pi^2}\, A \left( \mbox{div} - \ln{A} \right),
\label{integrala}
\ee
where $A$ is the mass squared of the scalar particle in the loop.
In order to compute the type (b) and type (c) Feynman diagrams
we need first to introduce a Feynman parameter $x$,
which is later integrated over from $x=0$ to $x=1$.
For type (b) diagrams we have
\ba
\mu^{4-d} \int \frac{\mbox{d}^d k}{\left( 2 \pi \right)^d}\,
\int_0^1 \mbox{d} x\,
\frac{4\, k^\mu k^\nu}{\left[ k^2 - A x - B \left( 1 - x \right)
+ i \varepsilon \right]^2}
&=& \frac{i g^{\mu \nu}}{16 \pi^2}
\left[
A \left( \mbox{div} - \ln{A} \right)
\right. \no & & \left.
+ B \left( \mbox{div} - \ln{B} \right)
+ F \left( A, B \right) \right],
\hspace*{6mm}
\label{integralc}
\ea
where $A$ and $B$ are the masses squared of the scalars in the loop,
and the four-momentum $q^\mu$ of the external gauge-boson line
is taken to obey $q^2 = 0$.
Notice the presence of terms
of the form $A \left( \mbox{div} - \ln{A} \right)$
in both diagrams of types (a) and (b);
we shall soon see that those terms
cancel out in the computation of $\Delta \rho$,
leaving only the $F$ functions from the type (b) diagrams.
For type (c) diagrams the relevant integral is
\ba
\mu^{4-d} \int \frac{\mbox{d}^d k}{\left( 2 \pi \right)^d}\,
\int_0^1 \mbox{d} x\,
\frac{g^{\mu \nu}}{\left[ k^2 - A x - B \left( 1 - x \right)
+ i \varepsilon \right]^2}
&=&
\frac{i g^{\mu \nu}}{16 \pi^2}\, \frac{1}{A} \left[
A \left( \mbox{div} - \ln{A} \right)
- \frac{A + B}{2}
\right. \no & & \left.
+ F \left( A, B \right) \right].
\label{integrald}
\ea
This integral is symmetric under the interchange of $A$ and $B$;
equation~(\ref{integrald}) presents a seemingly asymmetric form,
but it is in fact symmetric.
The reason for expressing the integral in this way is that,
due to cancellations,
only the terms $F \left( A, B \right)$ survive in the end.

\subsection{The contributions to $\Delta \rho$
from diagrams of types (a) and (b)}

Using~(\ref{int3}) and~(\ref{integrala}),
we see that the contribution to $A_{WW} \left( q^2 \right)$
of type (a) Feynman diagrams with charged scalars in the loop is
\be
A^{(a)}_{WW} \left( q^2 \right) = - \frac{g^2}{32 \pi^2}\,
\sum_{a=1}^n \left( U^\dagger U \right)_{aa}
m_a^2 \left( \mbox{div} - \ln{m_a^2} \right).
\label{aww}
\ee
In the same way,
using~(\ref{int4}),
\be
A^{(a)}_{ZZ} \left( q^2 \right) = - \frac{g^2}{32 \pi^2 c_W^2}\,
\sum_{a=1}^n \left[ 4 s_W^4 + \left( 1 - 4 s_W^2 \right)
\left( U^\dagger U \right)_{aa} \right]
m_a^2 \left( \mbox{div} - \ln{m_a^2} \right).
\label{azz}
\ee

Proceeding to the type (b) Feynman diagrams,
from~(\ref{int6}) and~(\ref{integralc}) we find that
\begin{subequations}
\ba
A^{(b)}_{WW} \left( 0 \right) &=&
\frac{g^2}{64 \pi^2}\, \sum_{a=1}^n \sum_{b=1}^m
\left( U^\dagger V \right)_{ab} \left( V^\dagger U \right)_{ba}
\left[
m_a^2 \left( \mbox{div} - \ln{m_a^2} \right)
\right. \no & & \left.
+ \mu_b^2 \left( \mbox{div} - \ln{\mu_b^2} \right)
+ F \left( m_a^2, \mu_b^2 \right) \right]
\no &=&
\frac{g^2}{64 \pi^2} \left[
2 \sum_{a=1}^n \left( U^\dagger U \right)_{aa}
m_a^2 \left( \mbox{div} - \ln{m_a^2} \right)
\right. \label{linha1} \\ & &
+ \sum_{b=1}^m \left( V^\dagger V \right)_{bb}
\mu_b^2 \left( \mbox{div} - \ln{\mu_b^2} \right)
\label{linha2} \\ & & \left.
+ \sum_{a=1}^n \sum_{b=1}^m
\left| \left( U^\dagger V \right)_{ab} \right|^2
F \left( m_a^2, \mu_b^2 \right) \right].
\label{linha3}
\ea
\end{subequations}
We have used 
\be
\sum_{a=1}^n 
\left( U^\dagger V \right)_{ab} \left( V^\dagger U \right)_{ba}
= \left( V^\dagger V \right)_{bb},
\ee
which follows from the unitarity of $\tilde U$,
i.e.~from~\cite{grimus}
\be
U U^\dagger = 1_{n_d \times n_d}.
\ee
We have also used
\be
\sum_{b=1}^m 
\left( U^\dagger V \right)_{ab} \left( V^\dagger U \right)_{ba}
= 2 \left( U^\dagger U \right)_{aa},
\ee
which follows from the orthogonality of $\tilde V$,
i.e.~from~\cite{grimus}
\be
\begin{array}{rcccl}
\mbox{Re}\, V\ \mbox{Re} V^T
&=& \mbox{Im}\, V\ \mbox{Im} V^T
&=& 1_{n_d \times n_d},
\\*[1mm]
\mbox{Re}\, V\ \mbox{Im} V^T
&=& \mbox{Im}\, V\ \mbox{Re} V^T
&=& 0_{n_d \times n_d}.
\end{array}
\label{ort3}
\ee
Considering now the self-energy of the $Z^0$ boson,
we find
\begin{subequations}
\ba
A_{ZZ}^{(b)} \left( 0 \right) &=&
\frac{g^2}{64 \pi^2 c_W^2} \left\{
\sum_{a, a^\prime = 1}^n
\left[ 2 s_W^2 \delta_{a a^\prime}
- \left( U^\dagger U \right)_{a^\prime a} \right]
\left[ 2 s_W^2 \delta_{a a^\prime}
- \left( U^\dagger U \right)_{a a^\prime} \right]
\right. \no & & \times \left[
m_a^2 \left( \mbox{div} - \ln{m_a^2} \right)
+ m_{a^\prime}^2 \left( \mbox{div} - \ln{m_{a^\prime}^2} \right)
+ F \left( m_a^2, m_{a^\prime}^2 \right) \right]
\no & &
+ \sum_{b=1}^{m-1} \sum_{b^\prime = b+1}^m
\left[ \mbox{Im} \left( V^\dagger V \right)_{b b^\prime} \right]^2
\no & & \left. \times
\left[
\mu_b^2 \left( \mbox{div} - \ln{\mu_b^2} \right)
+ \mu_{b^\prime}^2 \left( \mbox{div} - \ln{\mu_{b^\prime}^2} \right)
+ F \left( \mu_b^2, \mu_{b^\prime}^2 \right) \right]
\right\}
\no &=&
\frac{g^2}{64 \pi^2 c_W^2} \left\{
2\, \sum_{a=1}^{n-1} \sum_{a^\prime = a+1}^n
\left| \left( U^\dagger U \right)_{a a^\prime} \right|^2
F \left( m_a^2, m_{a^\prime}^2 \right)
\right. \label{linha4} \\ & &
+ 2\, \sum_{a=1}^n \left[ 4 s_W^4 + \left( 1 - 4 s_W^2 \right)
 \left( U^\dagger U \right)_{aa} \right]
m_a^2 \left( \mbox{div} - \ln{m_a^2} \right)
\label{linha5} \\ & &
+ \sum_{b=1}^{m-1} \sum_{b^\prime = b+1}^m
\left[ \mbox{Im} \left( V^\dagger V \right)_{b b^\prime} \right]^2
F \left( \mu_b^2, \mu_{b^\prime}^2 \right)
\label{linha6} \\ & & \left.
+ \sum_{b=1}^m \left( V^\dagger V \right)_{bb}
\mu_b^2 \left( \mbox{div} - \ln{\mu_b^2} \right)
\right\}.
\label{linha7}
\ea
\end{subequations}
We have used
\be
\sum_{b^\prime = 1}^m \left[ \mbox{Im}
\left( V^\dagger V \right)_{b b^\prime} \right]^2
= \left( V^\dagger V \right)_{bb},
\ee
which follows from equations~(\ref{ort3}).

Putting everything together,
we see that
\begin{description}
\item the $A^{(a)}_{WW} \left( q^2 \right)$ of equation~(\ref{aww})
cancels out the line~(\ref{linha1})
of $A^{(b)}_{WW} \left( 0 \right)$;
\item the $A^{(a)}_{ZZ} \left( q^2 \right)$ of equation~(\ref{azz})
cancels out the line~(\ref{linha5})
of $A^{(b)}_{ZZ} \left( 0 \right)$;
\item the line~(\ref{linha2}) of $A^{(b)}_{WW} \left( 0 \right)$
cancels out the line~(\ref{linha7}) of $A^{(b)}_{ZZ} \left( 0 \right)$
in the subtraction $A_{WW} - c_W^2 A_{ZZ}$.
\end{description}
In this way we finally obtain
\begin{subequations}
\label{l}
\ba
A^{(a+b)}_{WW} \left( 0 \right)
- c_W^2 A^{(a+b)}_{ZZ} \left( 0 \right)
&=& \frac{g^2}{64 \pi^2} \left\{
\sum_{a=1}^n \sum_{b=1}^m
\left| \left( U^\dagger V \right)_{ab} \right|^2
F \left( m_a^2, \mu_b^2 \right)
\right. \label{l1} \\ & &
- 2\, \sum_{a=1}^{n-1} \sum_{a^\prime = a+1}^n
\left| \left( U^\dagger U \right)_{a a^\prime} \right|^2
F \left( m_a^2, m_{a^\prime}^2 \right)
\label{l2} \\ & & \left.
- \sum_{b=1}^{m-1} \sum_{b^\prime = b+1}^m
\left[ \mbox{Im} \left( V^\dagger V \right)_{b b^\prime} \right]^2
F \left( \mu_b^2, \mu_{b^\prime}^2 \right)
\right\}. \hspace*{3mm}
\label{l3}
\ea
\end{subequations}
The positive term~(\ref{l1}) originates from $A^{(b)}_{WW}$
while the negative terms~(\ref{l2}) and~(\ref{l3})
come from $A^{(b)}_{ZZ}$.

If there are in the electroweak theory any scalar $SU(2)$ singlets
with electric charges other than $0$ or $\pm 1$,
then the relevant terms in the Lagrangian
are those in equation~(\ref{Lag}).
The term~(\ref{Lag2}) generates a type (a) Feynman diagram
which exactly cancels the type (b) Feynman diagram generated
by the term~(\ref{Lag1}).\footnote{This cancellation is analogous
to the one between equation~(\ref{azz}) and line~(\ref{linha5}).}
Thus,
scalar $SU(2)$ singlets
with electric charge different from $0$ and $\pm 1$
do not affect $\Delta \rho$ at all.

The sums in equation~(\ref{l}) include contributions
from the Goldstone bosons $G^\pm = S_1^\pm$ and $G^0 = S^0_1$.
These Goldstone bosons have unphysical masses $m_1$ and $\mu_1$,
respectively,
which are arbitrary in a 't~Hooft gauge.
The terms which depend on those masses are,
explicitly,
\begin{subequations}
\label{t}
\ba
& & \left| \left( U^\dagger V \right)_{11} \right|^2
F \left( m_1^2, \mu_1^2 \right)
\label{t1} \\
& & + \sum_{b=2}^m
\left| \left( U^\dagger V \right)_{1b} \right|^2
F \left( m_1^2, \mu_b^2 \right)
\label{t2} \\
& & + \sum_{a=2}^n
\left| \left( U^\dagger V \right)_{a1} \right|^2
F \left( m_a^2, \mu_1^2 \right)
\label{t3} \\
& & - 2\, \sum_{a=2}^n
\left| \left( U^\dagger U \right)_{1a} \right|^2
F \left( m_1^2, m_a^2 \right)
\label{t4} \\
& & - \sum_{b=2}^m
\left[ \mbox{Im} \left( V^\dagger V \right)_{1b} \right]^2
F \left( \mu_1^2, \mu_b^2 \right).
\label{t5}
\ea
\end{subequations}
One may eliminate some of these terms by
using equations~(\ref{uk1}) and~(\ref{vk1}).
Indeed,
$\left( U^\dagger U \right)_{1a}
= - \left( T^\dagger T \right)_{1a}
= 0$ when $a \neq 1$,
because $T_{j1} = 0$ for any $j$;
also,
$\left( U^\dagger V \right)_{a1} = i \left( U^\dagger U \right)_{a1} = 0$
for $a \neq 1$.
Therefore,
the terms~(\ref{t3}) and~(\ref{t4}) vanish.
In the term~(\ref{t1}),
$\left( U^\dagger V \right)_{11} = i$.
In the term~(\ref{t2}) one may write
\be
\left( U^\dagger V \right)_{1b} = i \left( V^\dagger V \right)_{1b}
= - \mbox{Im} \left( V^\dagger V \right)_{1b}\ \Leftarrow b \neq 1,
\ee
since $\mbox{Re} \left( V^\dagger V \right)_{1b}
= \left( \mbox{Re}\, V^T\ \mbox{Re}\, V
+ \mbox{Im}\, V^T\ \mbox{Im}\, V \right)_{1b}
= - \left( R^T R \right)_{1b} = 0$.
In this way,
the terms~(\ref{t}) are reduced to
\be
F \left( m_1^2, \mu_1^2 \right) + \sum_{b=2}^m
\left[ \mbox{Im} \left( V^\dagger V \right)_{1b} \right]^2
\left[ F \left( m_1^2, \mu_b^2 \right)
- F \left( \mu_1^2, \mu_b^2 \right) \right].
\label{tocancel}
\ee
The term $F \left( m_1^2, \mu_1^2 \right)$ is independent
of the number of scalar doublets and singlets,
hence it is eliminated when one subtracts the SM result
from the Multi-Higgs-doublet-model one.
The other terms in the expression~(\ref{tocancel})
are cancelled out by the diagrams of type~(c),
as we shall see next.

\subsection{The contributions to $\Delta \rho$
from diagrams of type (c)}

To compensate for the unphysical masses of the Goldstone bosons,
the propagators of gauge bosons $W^\pm$ and $Z^0$
with four-momentum $k^\mu$ are,
in a 't~Hooft gauge,
\ba
& & - \frac{k_\mu k_\nu}{m_W^2}\, \frac{i}{k^2 - m_1^2}
+ \left( - g_{\mu \nu} + \frac{k_\mu k_\nu}{m_W^2} \right)
\frac{i}{k^2 - m_W^2},
\label{Wprop} \\
& & - \frac{k_\mu k_\nu}{m_Z^2}\, \frac{i}{k^2 - \mu_1^2}
+ \left( - g_{\mu \nu} + \frac{k_\mu k_\nu}{m_Z^2} \right)
\frac{i}{k^2 - m_Z^2},
\label{Zprop}
\ea
respectively,
i.e.~they contain a piece with a pole
on the unphysical masses squared $m_1^2$ and $\mu_1^2$,
respectively.

Using these propagators to compute the type (c) Feynman diagrams,
one obtains
\ba
A^{(c)}_{WW} \left( 0 \right) &=&
\frac{g^2}{64 \pi^2}\,
\sum_{b=2}^m \left[ \mbox{Im} \left( V^\dagger V \right)_{1b} \right]^2
\left[ - m_1^2 \left( \mbox{div} - \ln{m_1^2} \right)
- 3 m_W^2 \left( \mbox{div} - \ln{m_W^2} \right)
\right. \no & & \left.
+ 2 \left( m_W^2 + \mu_b^2 \right)
- F \left( m_1^2, \mu_b^2 \right)
- 3 F \left( m_W^2, \mu_b^2 \right) \right],
\\
A^{(c)}_{ZZ} \left( 0 \right) &=&
\frac{g^2}{64 \pi^2 c_W^2}\,
\sum_{b=2}^m \left[ \mbox{Im} \left( V^\dagger V \right)_{1b} \right]^2
\left[ - \mu_1^2 \left( \mbox{div} - \ln{\mu_1^2} \right)
- 3 m_Z^2 \left( \mbox{div} - \ln{m_Z^2} \right)
\right. \no & & \left.
+ 2 \left( m_Z^2 + \mu_b^2 \right)
- F \left( \mu_1^2, \mu_b^2 \right)
- 3 F \left( m_Z^2, \mu_b^2 \right) \right].
\ea
The factors $3$ originate in a partial cancellation
between the contributions from the pieces
$- g_{\mu \nu}$ and $k_\mu k_\nu / m_V^2$
in the propagator of the gauge boson $V$,
the former contribution being four times larger than,
and with opposite sign relative to,
the latter one,
cf.~equations~(\ref{integralc}) and~(\ref{integrald}).
Performing the subtraction relevant for $\Delta \rho$,
one obtains
\begin{subequations}
\ba
A^{(c)}_{WW} \left( 0 \right)
- c_W^2 A^{(c)}_{ZZ} \left( 0 \right)
&=& \frac{g^2}{64 \pi^2}\,
\sum_{b=2}^m \left[ \mbox{Im} \left( V^\dagger V \right)_{1b} \right]^2
\no & & \times \left[
- m_1^2 \left( \mbox{div} - \ln{m_1^2} \right)
+ \mu_1^2 \left( \mbox{div} - \ln{\mu_1^2} \right)
\right. \label{l11} \\ & &
- 3 m_W^2 \left( \mbox{div} - \ln{m_W^2} \right)
+ 3 m_Z^2 \left( \mbox{div} - \ln{m_Z^2} \right) \hspace*{4mm}
\label{l12} \\ & &
+ 2 \left( m_W^2 - m_Z^2 \right)
\label{l13} \\ & &
- F \left( m_1^2, \mu_b^2 \right)
+ F \left( \mu_1^2, \mu_b^2 \right)
\label{l14} \\ & & \left.
- 3 F \left( m_W^2, \mu_b^2 \right)
+ 3 F \left( m_Z^2, \mu_b^2 \right) \right].
\label{drho}
\ea
\end{subequations}
The terms~(\ref{l11})--(\ref{l13})
are independent of the number of scalar doublets.
They disappear when one subtracts the Standard-Model result
from the multi-Higgs-doublet-model one,
since
\be
\sum_{b=2}^m \left[ \mbox{Im} \left( V^\dagger V \right)_{1b} \right]^2
= \left( V^\dagger V \right)_{11} = 1.
\ee
The terms~(\ref{l14}),
which involve the masses of the Goldstone bosons,
cancel out the terms in~(\ref{tocancel}),
except the first one,
which is cancelled by the subtraction of the SM result.

We have thus finished the derivation of equation~(\ref{final})
for $\Delta \rho$.

\section{The 2HDM and the Zee model}

In this section we give,
as examples of the application of our general formulae,
the expressions for $\Delta \rho$
in the 2HDM and also in the model of Zee~\cite{zee}
for the radiative generation of neutrino masses,
which has one singly charged $SU(2)$ singlet
together with the two doublets.

In the study of the 2HDM
it is convenient to use the so-called ``Higgs basis,''
in which only the first Higgs doublet has a vacuum expectation value.
In that basis,
\be
\phi_1 = \left( \begin{array}{c}
G^+ \\ \left( v + H + i G^0 \right) \left/ \sqrt{2} \right.
\end{array} \right),
\quad
\phi_2 = \left( \begin{array}{c}
S_2^+ \\ \left( R + i I \right) \left/ \sqrt{2} \right.
\end{array} \right).
\label{doublets}
\ee
Here,
$G^+ \equiv S_1^+$ and $G^0 \equiv S_1^0$ are the Goldstone bosons,
while $S_2^+$ is the physical charged scalar,
which has mass $m_2$.
Thus,
the matrix $U$,
which connects the charged components of $\phi_1$ and $\phi_2$
to the eigenstates of mass,
is in the Higgs basis of the 2HDM equal to the unit matrix.
On the other hand,
$H$,
$R$ and $I$,
which are real fields,
must be rotated through a $3 \times 3$ orthogonal matrix $O$
to obtain the three physical neutral fields $S^0_{2,3,4}$:
\be
\left( \begin{array}{c} H \\ R \\ I \end{array} \right)
= O \left( \begin{array}{c} S_2^0 \\ S_3^0 \\ S_4^0 \end{array} \right).
\ee
Without lack of generality we choose $\det{O} = +1$.
Thus,
the $2 \times 4$ matrix $V$,
defined through
\be
\left( \begin{array}{c} H + i G^0 \\ R + i I \end{array} \right)
= V
\left( \begin{array}{c} G^0 \\ S_2^0 \\ S_3^0 \\ S_4^0 \end{array} \right),
\ee
is
\be
V = \left( \begin{array}{cccc} i & O_{11} & O_{12} & O_{13} \\
0 & O_{21} + i O_{31} & O_{22} + i O_{32} & O_{23} + i O_{33}
\end{array} \right).
\label{vthdm}
\ee
Therefore,
\be
V^\dagger V = \left( \begin{array}{cccc}
1 & - i O_{11} & - i O_{12} & - i O_{13} \\
i O_{11} & 1 & i O_{13} & - i O_{12} \\
i O_{12} & - i O_{13} & 1 & i O_{11} \\
i O_{13} & i O_{12} & - i O_{11} & 1 \end{array} \right).
\label{vvthdm}
\ee
The value of $\Delta \rho$ in the 2HDM is therefore,
using our formula in equation~(\ref{finalMHDM}),
\ba
\Delta \rho &=& \frac{g^2}{64 \pi^2 m_W^2}
\left\{
\sum_{b=2}^4
\left( 1 - O_{1\, b-1}^2 \right)
F \left( m_2^2, \mu_b^2 \right)
\right. \no & &
- O_{13}^2 F \left( \mu_2^2, \mu_3^2 \right)
- O_{12}^2 F \left( \mu_2^2, \mu_4^2 \right)
- O_{11}^2 F \left( \mu_3^2, \mu_4^2 \right)
\no & & \left.
+ 3 \sum_{b=2}^4 O_{1\, b-1}^2 \left[ F \left( m_Z^2, \mu_b^2 \right)
- F \left( m_W^2, \mu_b^2 \right)
- F \left( m_Z^2, m_h^2 \right)
+ F \left( m_W^2, m_h^2 \right) \right] \right\}, \hspace*{7mm}
\label{thdm}
\ea
where $\mu_{2,3,4}$ denote the the masses of $S^0_{2,3,4}$,
respectively,
while $m_h$ is the mass of the Higgs boson of the SM.
Equation~(\ref{thdm}) reproduces,
in a somewhat simplified form,
the result for $\Delta \rho$ in the 2HDM previously given in~\cite{osland}.

A special case of the 2HDM is the model
with one ``dark'' scalar doublet. This means that a second doublet is added
to the SM,
but that doublet has no VEV
and it does not mix with the standard Higgs doublet~\cite{Ma:2006km}.
We should then identify $H$ with the usual Higgs particle.
Thus,
$O_{11} = 1$ and $\mu_2 = m_h$.
Equation~(\ref{thdm})
then simplifies to~\cite{Lavoura:1993mz,Barbieri:2006dq}
\be
\Delta \rho = \frac{g^2}{64 \pi^2 m_W^2} \left[
\sum_{b=3}^4 F \left( m_2^2, \mu_b^2 \right)
- F \left( \mu_3^2, \mu_4^2 \right) \right].
\ee
This quantity is small if the three masses $m_2$,
$\mu_3$ and $\mu_4$ are close together.
Notice that in this case of a ``dark'' scalar doublet
there are no vector--vector--scalar couplings
involving the additional doublet,
hence $\Delta \rho$ stems exclusively
from type~(a) and type~(b) Feynman diagrams.

In the model of Zee there is,
besides the two scalar $SU(2)$ doublets
\be
\phi_1 = \left( \begin{array}{c}
G^+ \\ \left( v + H + i G^0 \right) \left/ \sqrt{2} \right.
\end{array} \right),
\quad
\phi_2 = \left( \begin{array}{c}
H^+ \\ \left( R + i I \right) \left/ \sqrt{2} \right.
\end{array} \right),
\ee
also one scalar $SU(2)$ singlet $\chi^+$ with unit electric charge.
Therefore there is a $2 \times 2$ unitary matrix $K$ such that
\be
\left( \begin{array}{c} H^+ \\ \chi^+ \end{array} \right)
= K \left( \begin{array}{c} S_2^+ \\ S_3^+ \end{array} \right),
\ee
where $S_2^+$ and $S_3^+$ are the physical charged scalars,
which have masses $m_2$ and $m_3$,
respectively.
So,
now the matrix $U$ of equation~(\ref{Umatrix}) is
\be
U = \left( \begin{array}{ccc} 1 & 0 & 0 \\
0 & K_{11} & K_{12} \end{array} \right),
\ee
so that
\be
U^\dagger U = \left( \begin{array}{ccc}
1 & 0 & 0 \\
0 & \left| K_{11} \right|^2 & K_{11}^\ast K_{12} \\
0 & K_{11} K_{12}^\ast & \left| K_{12} \right|^2
\end{array} \right).
\ee
Equations~(\ref{vthdm}) and (\ref{vvthdm}) retain their validity,
and
\be
U^\dagger V = \left( \begin{array}{cccc}
i & O_{11} & O_{12} & O_{13} \\
0 &
K_{11}^* \left( O_{21} + i O_{31} \right) &
K_{11}^* \left( O_{22} + i O_{32} \right) &
K_{11}^* \left( O_{23} + i O_{33} \right) \\
0 &
K_{12}^* \left( O_{21} + i O_{31} \right) &
K_{12}^* \left( O_{22} + i O_{32} \right) &
K_{12}^* \left( O_{23} + i O_{33} \right)
\end{array} \right).
\ee
Therefore,
using our general formula~(\ref{final}) for $\Delta \rho$,
we see that,
in the model of Zee,
\ba
\Delta \rho &=& \frac{g^2}{64 \pi^2 m_W^2}
\left\{
\sum_{b=2}^4 \left( 1 - O_{1\, b-1}^2 \right)
\sum_{a=2}^3
\left| K_{1\, a-1} \right|^2 F \left( m_a^2, \mu_b^2 \right)
\right. \nonumber \\*[1mm] & &
- 2 \left| K_{11} K_{12} \right|^2 F \left( m_2^2, m_3^2 \right)
\nonumber \\*[2mm] & &
- O_{13}^2 F \left( \mu_2^2, \mu_3^2 \right)
- O_{12}^2 F \left( \mu_2^2, \mu_4^2 \right)
- O_{11}^2 F \left( \mu_3^2, \mu_4^2 \right)
\no & & \left.
+ 3 \sum_{b=2}^4 O_{1\, b-1}^2 \left[ F \left( m_Z^2, \mu_b^2 \right)
- F \left( m_W^2, \mu_b^2 \right)
- F \left( m_Z^2, m_h^2 \right)
+ F \left( m_W^2, m_h^2 \right) \right] \right\}. \hspace*{7mm}
\label{zee}
\ea

\section{Summary}

In this paper we have derived the formula
for the parameter $\Delta \rho$,
as defined in equation~(\ref{delta-rho0}),
in an extension of the Standard Model characterized by
an arbitrary number of scalar $SU(2)$ doublets
(with hypercharge $\pm 1/2$)
and singlets
(with arbitrary hypercharges).
Our formalism is completely general, 
using only
the masses of the scalars and their mixing matrices,
which ensures that our formulae are always applicable.
The computation has been carried out in a general $R_\xi$ gauge,
thereby demonstrating that the final result
is independent of the masses of the unphysical scalars.
We have also explicitly demonstrated that all infinities cancel out
in the final result for $\Delta \rho$. 
In order to ease the consultation of this paper,
the formulae for $\Delta \rho$ given in Section~2  
have been completely separated
from their derivation presented in Section~3.
Our results can be applied either to check the viability of a model
or to constrain its parameter space,
by comparing the $\Delta \rho$,
calculated in that model,
with numerical bounds on $\Delta \rho$
obtained from a fit to precision data---for instance,
the bound~(\ref{value}) found in~\cite{delta-rho-th}.
As an illustration of our general formulae,
in Section~4 we have worked out the specific cases
of the two-Higgs-doublet model,
with and without one extra charged scalar singlet.

\paragraph{Acknowledgements:}
W.G. thanks S.~Dittmaier,
W.~Hollik,
M.~Krawczyk and H.~Neu\-feld for helpful discussions. 
The work of L.L.\ was supported by the Portuguese
\textit{Funda\c c\~ao para a Ci\^encia e a Tecnologia}
through the project U777--Plurianual.
W.G.\ and L.L.\ acknowledge support from EU under the
MRTN-CT-2006-035505 network programme.

\end{document}